\documentclass{nature_arx}
\usepackage[format=plain, font=bf, justification=justified, singlelinecheck=false]{caption}

\usepackage{graphicx}

\providecommand{\e}[1]{\ensuremath{\times 10^{#1}}}

\title{Extremely metal-poor stars from the cosmic dawn in the bulge of the Milky Way}

\author{L. M. Howes$^{1}$, A. R. Casey$^{2}$, M. Asplund$^{1}$, S. C. Keller$^{1}$, D. Yong$^{1}$, D. M. Nataf$^{1}$, R. Poleski$^{3,4}$, K. Lind$^{5}$, C. Kobayashi$^{1,6}$, C. I. Owen$^{1}$, M. Ness$^{7}$, M. S. Bessell$^{1}$, G. S. Da Costa$^{1}$, B. P. Schmidt$^{1}$, P. Tisserand$^{1,8}$, A. Udalski$^{3}$, M. K. Szyma{\'n}ski$^{3}$, I. Soszy{\'n}ski$^{3}$, G. Pietrzy{\'n}ski$^{3,9}$, K. Ulaczyk$^{3,10}$, {\L}. Wyrzykowski$^{3}$, P. Pietrukowicz$^{3}$, J. Skowron$^{3}$, S. Koz{\l}owski$^{3}$, \& P. Mr{\'o}z$^{3}$}

\begin{document}

\pagenumbering{gobble}

\maketitle

\begin{affiliations}
 \item Research School of Astronomy and Astrophysics, Australian National University, ACT 2601, Australia
 \item Institute of Astronomy, University of Cambridge, Madingley Road, Cambridge, CB3 0HA, United Kingdom
 \item Warsaw University Observatory, Al. Ujazdowskie 4, 00-478 Warszawa, Poland
 \item Department of Astronomy, Ohio State University, 140 W. 18th Ave., Columbus, OH 43210, USA
 \item Department of Physics and Astronomy, Division of Astronomy and Space Physics, Uppsala University, Box 516, SE-751 20 Uppsala, Sweden
 \item School of Physics, Astronomy and Mathematics, Centre for Astrophysics Research, University of Hertfordshire, College Lane, Hatfield AL10 9AB, United Kingdom
 \item Max-Planck-Institut f\"ur Astronomie, K\"onigstuhl 17, D-69117 Heidelberg, Germany
 \item Sorbonne Universit\'es, UPMC Univ Paris 6 et CNRS, UMR 7095, Institut d'Astrophysique de Paris, 98 bis bd Arago, 75014 Paris, France
 \item Universidad de Concepci{\'o}n, Departamento de Astronomia, Casilla 160-C, Concepci{\'o}n, Chile
 \item Department of Physics, University of Warwick, Gibbet Hill Road, Coventry, CV4 7AL, United Kingdom
\end{affiliations}

\begin{abstract}
The first stars are predicted to have formed within 200 million years after the Big Bang\cite{Bromm_2009}, initiating the cosmic dawn.
A true first star has not yet been discovered, although stars\cite{Christlieb_2002,Caffau_2011,Keller_2014} with tiny amounts of
elements heavier than helium (`metals')
have been found in the outer regions (`halo') of the Milky Way.  The first stars and their immediate successors should, however, preferentially be found today in
the central regions (`bulges') of galaxies, because they
formed in the largest over-densities that grew gravitationally with time 
\cite{Tumlinson_2009,Salvadori_2010}. The Milky Way bulge underwent a rapid chemical enrichment during the first 1-2 billion years \cite{Feltzing_1999}, leading to a dearth of early, metal-poor stars \cite{Garc_a_P_rez_2013,Howes_2014}. Here we
report the observations of extremely metal-poor stars in the Milky Way bulge, 
including one star with an iron abundance about 10,000 times lower than the solar value without noticeable carbon enhancement. 
We confirm that most of the metal-poor bulge stars are on tight orbits around the Galactic Centre,
rather than being halo stars passing through the bulge, as expected for stars formed at redshifts greater than 15.
Their chemical compositions are in general similar to typical halo stars of the same metallicity although intriguing
differences exist, including lower abundances of carbon.
\end{abstract}

\begin{figure}
\centering
\includegraphics[width=0.99\columnwidth]{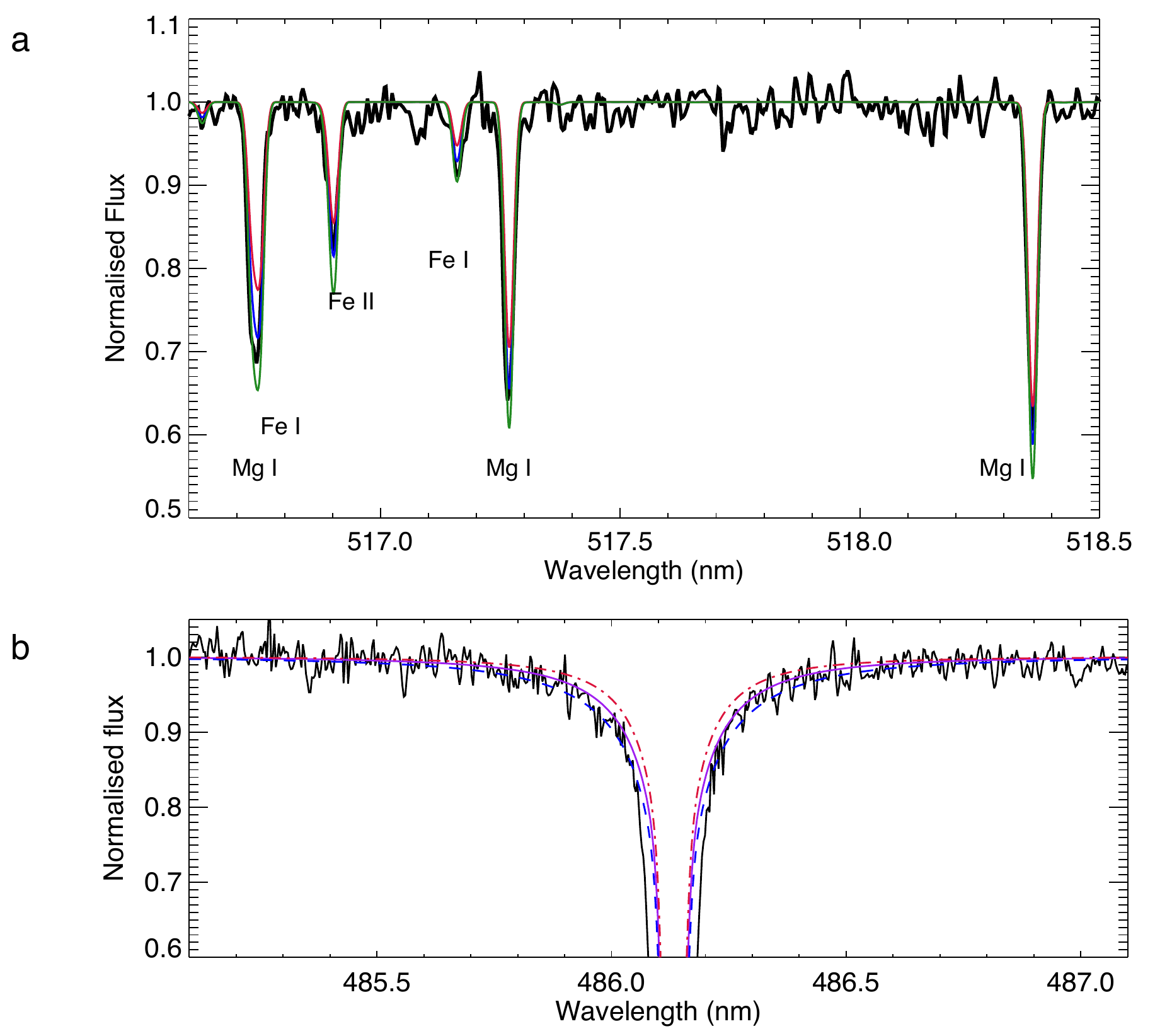}
\caption{\footnotesize
Extracts of the spectrum of the lowest metallicity star in our sample. \textbf{a:} A section of the spectrum of SMSS J181609.62-333218.7 (black line), the most metal-poor bulge star known. In blue is the predicted spectrum with the inferred stellar parameters (effective temperature $T_{\rm eff} = 4809$\,K, $\log g = 1.93$ dex [cgs], [Fe/H]$=-3.94$, [Mg/Fe]$=0.20$), and the red and green lines show spectra with all abundances scaled to +/-\,0.15\,dex respectively. All three were created using the 1D local thermodynamic equilibrium (LTE) spectrum synthesis programme, \textsc{moog} \cite{2012ascl.soft02009S}. \textbf{b: } The H$\beta$ line of the same star, compared to three synthetic spectral line profiles \cite{Barklem_2000} computed with $T_{\rm eff} =4640$\,K (red, dash-dot), $4800$\,K (purple, continuous), and $4960$\,K (blue, dashed).
\label{fig:spectrum}}
\end{figure}

Stars with a low content of heavy elements have distinct spectral flux distributions, which are reflected in their colours. Using the photometric filter system on the SkyMapper telescope operated by the Australian National University, it is possible to identify metal-poor candidate stars \cite{Keller_2007} in the Galactic halo \cite{Keller_2014} and bulge \cite{Howes_2014}. We have observed $\sim14,000$ bulge stars preselected from SkyMapper photometry using the AAOmega spectrograph on the Anglo-Australian Telescope (AAT), which enables the acquisition of 400 simultaneous stellar spectra over a 2-degree field of view. More than 500 stars with an iron abundance less than 1/100th of the solar value have been identified, making our survey the first to successfully target metal-poor stars in the Milky Way bulge. Twenty-three of these stars, targeted as the most metal-poor ones on the basis of the intermediate resolution spectra (Extended Data Table \ref{table:photo}), were observed in June 2014 with the MIKE high-resolution spectrograph on the 6.5-m Magellan Clay telescope \cite{Bernstein_2003} to enable a comprehensive determination of their chemical compositions (Fig. \ref{fig:spectrum}).

The stars' effective temperatures were derived through fitting the observed hydrogen lines with theoretical spectra, while neutral and ionized iron lines provided measurements of the surface gravities and metallicities in the framework of 1D stellar atmosphere models \cite{Gustafsson_2008} and non-equilibrium spectral line formation \cite{Lind_2012} (Extended Data Table \ref{table:params}). All 23 stars were found to have [Fe/H]$\leq$$-2.3$, including nine stars with [Fe/H]$<$$-3$ (here [A/B] = $\log_{10}(\frac{N_{A}}{N_{B}})_{*} - \log_{10}(\frac{N_{A}}{N_{B}})_{\odot}$, where $\frac{N_{A}}{N_{B}}$ refers to the number ratio of atoms of elements A and B in the star ($_{*}$) and the Sun ($_{\odot}$)). 
The most metal-poor star, SMSS J181609.62-333218.7, has [Fe/H]$=-3.94 \pm 0.16$.
The abundances of an additional 22 elements were determined spectroscopically, including the $\alpha$-elements Mg, Si, Ca, and Ti, and the neutron capture elements Y, Zr, and Ba (Extended Data Tables \ref{table:abunds}, \ref{table:abunds2}, \ref{table:abunds3}). 

\begin{figure}
\centering
\includegraphics[width=0.7\columnwidth]{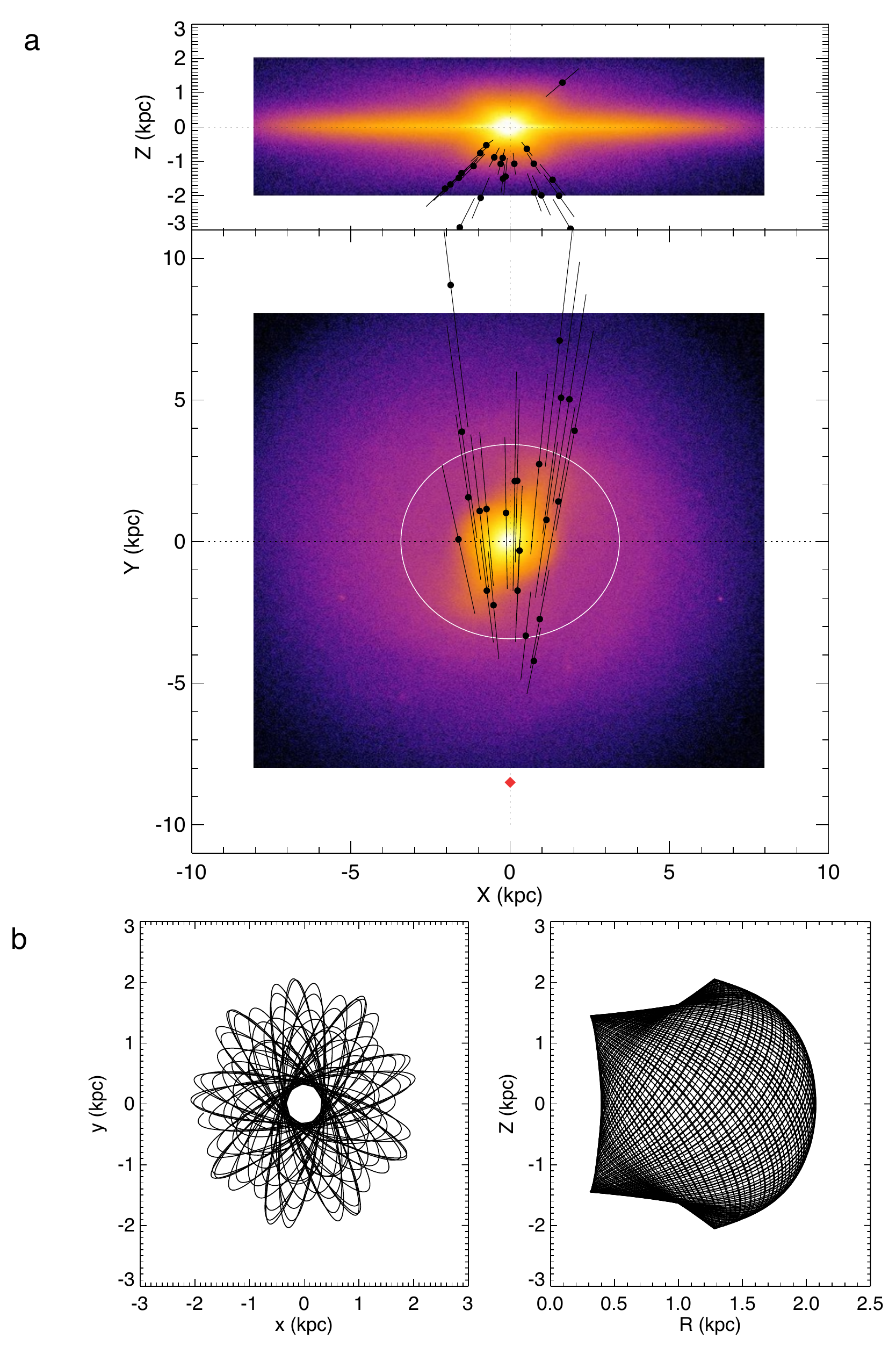}
\caption{\footnotesize
The Galactic positions and orbits of the 23 stars observed at high resolution. \textbf{a:} Surface density map of a model of the Galactic bulge projected onto the X-Z (top) and X-Y (bottom) planes \cite{Ness_2014}, where X, Y, and Z are Cartesian coordinates with the origin at the Galactic Centre and Z perpendicular to the plane of the Galaxy. Plotted over this *filled black circles) are the 23 stars of this study, with distance uncertainties shown as error bars, and a circle of radius 3.43\,kpc (white: the cutoff radius of the inner bulge determined from 2MASS data \cite{Robin_2012}). The position of the Sun is shown with a red diamond, at 8.5\,kpc from Galactic Centre.
\textbf{b:} Projections of the orbit of the lowest metallicity star, SMSS J181609.62-333218.7, both in the (R,Z) plane (right), where R is the radial direction, and in the plane of the orbit itself (left).
\label{fig:orbits}}
\end{figure}

To confirm their bulge membership, the distances and orbits of the stars have been determined.
Using the spectroscopic temperatures and surface gravities, and an assumed mass of $0.8\,{\mathrm M}_{\odot}$, distances were inferred, which in nearly all cases are consistent with them being located within the bulge (Fig. \ref{fig:orbits}). We have measured velocities for ten of our stars using observations taken by the OGLE-IV survey \cite{Udalski_2015}, from which orbits around the Galaxy have been determined in combination with their distances and velocities (Extended Data Table \ref{table:orbits}); the remaining stars fall outside the OGLE footprint while other sources of kinematic information are too uncertain to constrain the orbits sufficiently. Seven out of the ten stars with accurate kinematics are shown to have tightly bound orbits, placing them in the inner regions of the Milky Way (Fig. \ref{fig:orbits}). In particular, using a cutoff radius of 3.43\,kpc as the radius of the bulge component\cite{Robin_2012}, the most metal-poor star SMSS J181609.62-333218.7 has an orbit entirely contained within the bulge. Only two out of the ten stars are on much larger orbits, being merely halo stars currently passing through the bulge region. Extending these numbers to the whole sample, we can expect $\sim$ 14 of the 23 bulge stars analysed here to have orbits fully within the central regions of the Milky Way; with the imminent arrival of kinematic data from the Gaia satellite, accurate orbits for all of the bulge stars will be able to be determined.

The very first stars are predicted to have brought about the cosmic dawn by forming in the centres of the largest dark matter mini-halos, which subsequently accreted material to become the inner regions of the largest galaxies\cite{Greif_2012}. The typical redshift of formation for stars in the bulge with [Fe/H]$<$$-1$ is $z\approx 10$, in contrast to $z\approx 5$ for halo stars. Of the stars with [Fe/H]$<$$-3$, approximately $15\%$ are expected to have formed at $z>15$ (\cite{Tumlinson_2009,Salvadori_2010}). Of the ten stars with accurate orbit information, half of them have binding energies E$_{tot}$$<$$-8\e{-4}$\,km$^{2}$\,s$^{-2}$, which is consistent with a formation redshift of $z>15$ (\cite{Tumlinson_2009}). Low binding energies imply that the stars have been in the Galactic potential well for some time and it is very unlikely they have been accreted from a recent dwarf spheroidal merger.
Their low metallicities, orbits and binding energies make these stars prime candidates for being direct descendants of the very first stars, probing a cosmic epoch otherwise completely inaccessible currently.  
Direct age determinations of these ancient and extremely metal-poor bulge stars from comparison with stellar evolutionary tracks or radioactive U or Th dating are currently not possible but asteroseismic ages could possibly be inferred with the extended Kepler mission or future satellites.  

Given their extremely low metallicities and large formation redshifts, these stars are likely to have formed from gas polluted by ejecta from a single or at most a few supernovae of the first stellar generation.  A chemical composition analysis has been carried out to search for tell-tale nucleosynthetic signatures and possible differences from halo stars at the same metallicities. For most elements, the chemical compositions of the 23 bulge stars are consistent with typical halo stars, suggesting enrichment by similar supernovae in spite of the distinct environments and formation redshifts. Subtle differences do exist however, most notably in terms of the carbon abundances. None of the 23 stars have the large observed carbon enhancements that occur frequently in halo stars.  Applying evolutionary corrections to the surface carbon abundance to counter the mixing that occurs with material processed by H-burning through CNO-cycling at late stages of the stellar lifetime \cite{Placco_2014}, still only one of the stars would have had a natal [C/Fe]$>$$1$. In the halo, the percentage of stars that are carbon-enhanced increases dramatically at lower metallicities - from 27\% of stars with [Fe/H]$<$$-2$ up to 69\% with [Fe/H]$<$$-4$ (\cite{Placco_2014}).  
From the literature data on halo stars with similar iron abundances to our stars\cite{Placco_2014}, the probability of selecting at most one carbon-enhanced star out of 23 halo stars is only 0.2\%. Carbon-enhanced stars come in two varieties; those with and those without large excesses of neutron-capture elements. The former are most likely to have been formed by mass-transfer from a binary companion that underwent the asymptotic giant branch phase. Those carbon/enhanced stars with neutron capture excesses occur most frequently at metallicities of [Fe/H]$>$$-3$, whereas those without do not appear to have binary companions, and are more common at the very lowest metallicities. As none of our bulge stars are classified as having large abundances of neutron capture elements, the likelihood of finding one such carbon-enhanced star out of 23 is 7\% if the frequency is the same for the bulge as for the halo. A lower frequency of carbon-enhanced stars in the bulge relative to the halo is contrary to theoretical predictions; the expected dependence of the initial mass function on the cosmic microwave background\cite{Tumlinson_2007} would result in a greater number of carbon-enhanced stars near the centre of the Galaxy.

The most metal-poor bulge star, SMSS J181609.62-333218.7 is at least an order of magnitude more iron-deficient than previously found low-metallicity bulge stars  \cite{Howes_2014,Schlaufman_2014}. We have not been able to detect C in its spectrum, instead finding only an upper limit to the star's C abundance (Extended Data Figure 1). This makes the upper limit on its total metallicity, [Z/H]$\sim$$-3.8$ (total mass fraction of Z$\sim2.1$$\times10^{-6}$), where Z represents the sum of all metals, placing it amongst the four most metal-poor stars known, along with the halo star SDSS J102915+172927 (\cite{Caffau_2011}). The low C measured in both these stars fall below the predicted metallicity limit for formation of low-mass stars due to metal line cooling\cite{Frebel_2007}.

\begin{figure}
\centering
\includegraphics[width=0.85\columnwidth]{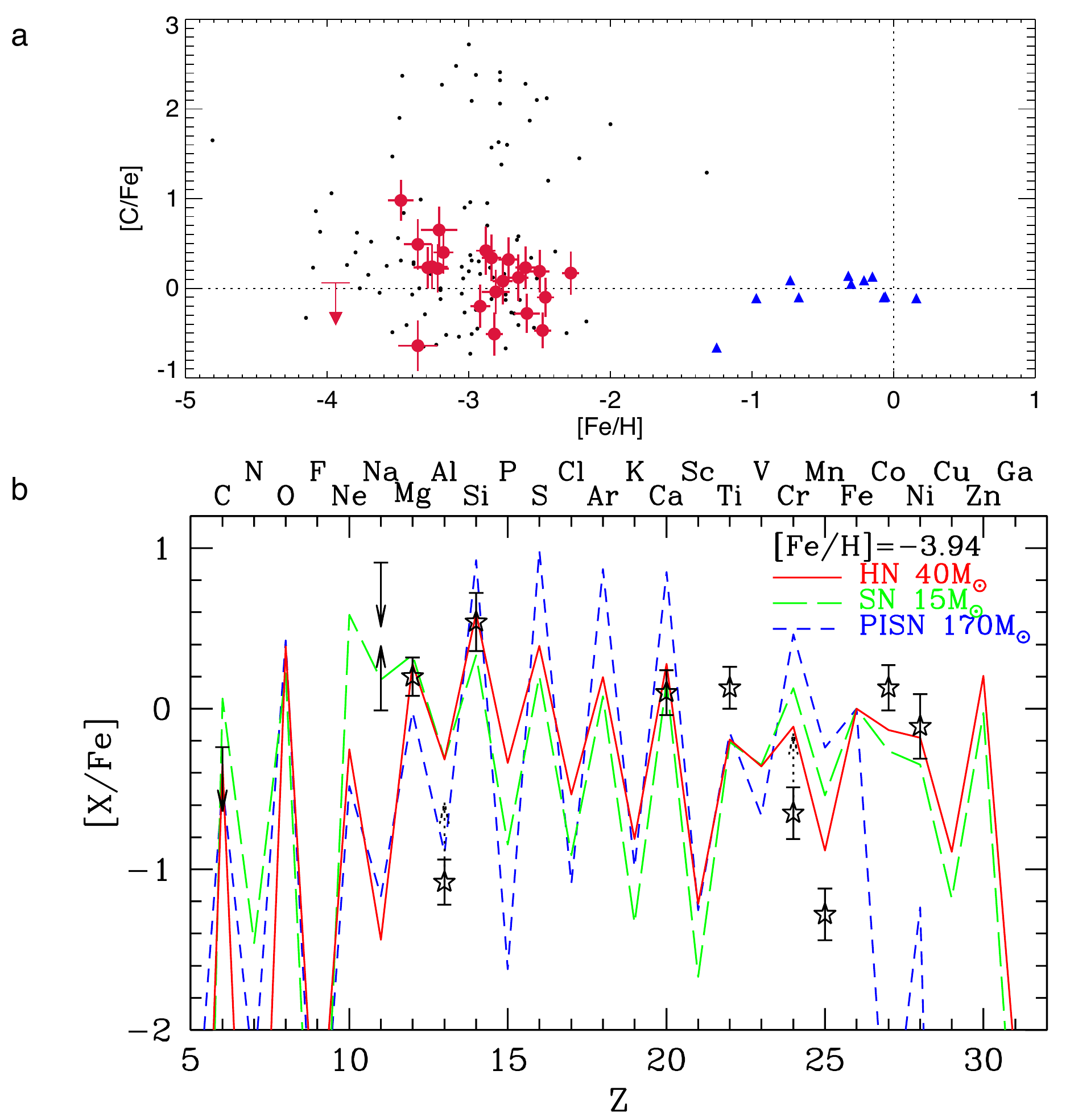}
\caption{
\footnotesize Chemical abundances of the 23 stars observed at high resolution. \textbf{a:} The abundance ratio of carbon versus iron ([C/Fe]), with respect to metallicity ([Fe/H]) measured in the observed stars (filled red circles, red arrow for an upper limit). The dotted lines represent the solar abundance. Also shown for comparison are literature metal-poor halo giants (small black dots \cite{Yong_2012}) and more metal-rich bulge stars (filled blue triangles \cite{Ryde_2010}). \textbf{b:} The chemical abundance pattern of SMSS J181609.62-333218.7, for elements X, where X is displayed at the top of the figure. Each determined abundance is shown as an open black star. These abundances are compared to three synthetic supernovae yields; a pair-instability supernova of $170\,{\mathrm M}_{\odot}$ (PISN; blue, short dash \cite{Umeda_2002}), a core-collapse supernova of $15\,{\mathrm M}_{\odot}$ (SN; green, long dash \cite{Kobayashi_2014}), and a hypernova of $40\,{\mathrm M}_{\odot}$ (HN; red, solid \cite{Kobayashi_2014}). Dotted arrows represent expected non-LTE corrections, solid arrows represent measurements where only an upper or lower limit was possible. The error bars in parts a and b are estimates of the uncertainties in our measurements, calculated as described in Methods.
\label{fig:abunds}}
\end{figure}

We have compared the detailed chemical abundance pattern of SMSS J181609.62-333218.7 to primordial supernovae yields \cite{Kobayashi_2014, Umeda_2002} (Fig. \ref{fig:abunds}). In particular, the low Mg and Ca abundance, but higher Si abundance, and the absence of a pronounced odd-even abundance pattern rule out the possibility of enrichment by a pair-instability supernova resulting from a primordial star of $140-250\,{\mathrm M}_{\odot}$. Low abundances of Cr and Mn and of $\alpha$-elements, combined with the higher abundance of Co, indicate that the polluting supernova was most likely to have been a primordial hypernova -- an extremely energetic kind of supernova releasing ten times the kinetic energy of regular core-collapse supernovae, possibly due to the forming black hole having larger angular momentum\cite{Nomoto_2003}. Good agreement is found for a $40\,{\mathrm M}_{\odot}$ hypernova; a more stringent Zn limit would further constrain the mass range. Unusual abundance ratios have been found in small numbers of stars in the halo -- 4\% of halo stars with low carbon abundances have chemical peculiarities in at least two elements \cite{Yong_2012} - but none so far appear to have been polluted by a $40\,{\mathrm M}_{\odot}$ hypernova. A low [$\alpha$/Fe] ratio ($0.14$\,dex) at such low [Fe/H] is consistent with an inhomogeneous enrichment from such supernovae\cite{Karlsson_2013}, while stars with higher [$\alpha$/Fe] formed from more well-mixed gas due to a longer time delay in forming the second generation of stars.

\subsection{Methods}

\subsubsection*{Observations}
Photometry of the Milky Way bulge was acquired for the EMBLA survey \cite{Howes_2014} during the commissioning period of the SkyMapper telescope in 2012 and 2013. Stars were selected from the photometry using a combination of the $g$, $i$, and $v$ bandpasses, designed to give a reliable metallicity indicator \cite{Keller_2007}.
\\
Spectroscopic follow-up observations took place during 2012-14, making use of the AAOmega+2dF multi-object spectrograph \cite{Sharp_2006} on the Anglo-Australian Telescope. With between 350 and 400 stars observed in each field, spectra of more than 14,000 bulge stars have been obtained. The gratings used have a spectral resolving power of 1,300 in the blue (370-580\,nm) and 10,000 in the red (840-885\,nm). The data were reduced using the standard 2dfdr pipeline. Stellar spectra were fitted using a generative model that simultaneously accounts for stellar parameters (by interpolating from the AMBRE grid \cite{de_Laverny_2012}), continuum, spectral resolution and radial velocity. 
\\
From the first two years of spectroscopic data, more than 50 stars were identified as having [Fe/H]$<-2.5$. The high-resolution spectroscopic data of 23 stars presented in this Letter are the result of observations using the MIKE spectrograph at the Magellan Clay telescope \cite{Bernstein_2003} on the 15-17 June 2014. All observations were taken using a slit width of 0.7", resulting in a resolving power of 35,000 in the blue and 31,000 in the red. The data were reduced using the CarPy data reduction pipeline \cite{carpy}, before they were normalised and summed together using the \textsc{smh} software \cite{2014arXiv1405.5968C}. The final spectra cover 330-890\,nm.

\subsubsection*{Parameter and Abundance Determination}
The stellar parameters (Extended Data Table \ref{table:params}) were calculated iteratively, using the original parameters from the low-resolution spectra as initial guesses. First, effective temperatures were derived by fitting the wings of the Balmer H$\alpha$ and H$\beta$ lines with a synthetic profile (Fig. \ref{fig:spectrum}). These profiles were created by linearly interpolating between a grid of synthetic spectra \cite{Barklem_2000}. The best lines were fit by a $\chi^{2}$-minimisation, using a weighted average of the two lines - weighting was double on the H$\beta$ line, due to predicted LTE effects being larger for H$\alpha$ (\cite{Barklem_2007}). The difference between the temperatures calculated for each line was on average only $26$\,K. The $\log{g}$, microturbulence $\xi_{t}$, and [Fe/H] were then derived for that temperature, by forcing the Fe\textsc{i} abundance to remain constant with respect to reduced equivalent width, and equilibrium between the Fe\textsc{i} and Fe\textsc{ii} abundances.  Fe\textsc{i} and Fe\textsc{ii} abundances were measured from the equivalent widths of a maximum of 66 Fe\textsc{i} lines and 24 Fe\textsc{ii} lines (in the case of the most metal-poor star, SMSS J181609.62-333218.7, these numbers are reduced to 10 Fe\textsc{i} lines and 4 Fe\textsc{ii} lines). Finally a non-LTE correction is applied to the Fe\textsc{i} abundance, calculated by taking the average of the line-by-line corrections \cite{Lind_2012}. This correction forces an offset between the Fe\textsc{i} and Fe\textsc{ii} abundances, thus replacing the initial equilibrium. This process is repeated until the parameters converge on a solution. Throughout we use the 1D MARCS model atmospheres \cite{Gustafsson_2008}, and a shortened version of the Gaia-ESO line list, with extra lines supplemented from \cite{Norris_2013} due to our wider wavelength coverage. The stellar abundances are referenced to the solar abundances of ref. \cite{Asplund_2009}. This analysis method was tested on seven halo stars from the literature \cite{Yong_2012}, and the offsets found were $T_{\rm eff}=+28K$, $\log{g}=-0.2$, and [Fe/H]$=-0.08$ (literature values - our values).
\\
The abundances were measured using the equivalent widths of atomic lines (that were all on the linear part of the curve-of-growth), except in the case of C (measured from the C-H molecular bands at 431.3\,nm and 432.3\,nm) and Ba (synthesised in order to account for hyperfine splitting). Non-LTE corrections were calculated for Li (\cite{Lind_2009}), Na (\cite{Lind_2011}), Mg, and Ca, and applied to the individual line abundances. The literature halo abundances of Mg and Ca (\cite{Yong_2012}) shown in Figure \ref{fig:abunds} have also had a NLTE correction applied, in order to ensure a fair comparison. 3$\sigma$ upper limits were derived for some elements in those stars where the lines were too weak to be detected (Extended Data Table \ref{table:abunds}). The abundance offsets compared to the literature values averaged 0.10$\pm0.19$ across those elements measured in common. Due to wavelengths covered in the SkyMapper metallicity filter, it is possible that stars with extremely high C abundances appeared to be more metal-rich, and so weren't selected. However, a similar study of metal-poor stars discovered in the halo with SkyMapper \cite{Jacobson_2015} found the fraction of C-enhanced stars was identical to that reported in previous surveys\cite{Lucatello_2006}. Furthermore, we followed up 14,000 stars with intermediate resolution spectra, and determined metallicities using those spectra. The majority of the stars observed had [Fe/H]$\approx$$-1.0$ and included some that had solar metallicities, so it is highly unlikely that we missed any C-enhanced EMP star in our selection. 
\\
The systematic uncertainties in the temperature determinations were estimated to be $\pm100$\,K, and the statistical uncertainties averaged $\pm125$\,K, so combined in quadrature we conclude the total uncertainty to be $\pm160$\,K. The $\xi_{t}$ uncertainties are estimated to be $\pm0.2$, mostly due to systematics. The standard errors of the individual line abundances of Fe\textsc{i} and Fe\textsc{ii} were combined in quadrature to evaluate the $\log{g}$ uncertainties. The differences between the [Fe/H] values when varying the temperature, surface gravity, and microturbulence by their respective errors were combined in quadrature with the standard error of the Fe\textsc{ii} lines to produce the [Fe/H] uncertainties. The individual abundance errors were also calculated using this method, using the standard error of the individual abundances for the lines of that particular element.
\\
For SMSS J181609.62-333218.7, which has [Fe/H]$=-3.94$, a measurement of Na was not possible, due to the 818.3\,nm and 819.4\,nm lines being too weak, and the Na D lines (588.9\,nm and 589.5\,nm) being partly blended with interstellar Na lines. We have derived a range of possible values for this star, taking the upper limit from the non-detection at 819.4\,nm, and the lower limit from fitting a Gaussian to the Na D lines, taking into account the interstellar Na.

\subsubsection*{Distances and Orbital Parameters}
Distances to the stars were calculated by comparing the absolute and apparent bolometric magnitudes. The absolute magnitudes were recovered from the relation
${\mathrm M}_{*}={\mathrm M}_{\odot}-2.5\log{\frac{{\mathrm L}_{*}}{{\mathrm L}_{\odot}}}$,
where the luminosities are calculated using
$\frac{{\mathrm L}_{*}}{{\mathrm L}_{\odot}}={\frac{4 \pi \sigma {\mathrm T}^{4}{\mathrm M}_{*}G}{10^{\log{g_{*}}}}}$,
taking ${\mathrm M}_{*}=0.8 \pm0.2{\mathrm M}_{\odot}$ for all stars. The apparent bolometric magnitudes are reconstructed from the 2MASS $JHK_{s}$ magnitudes (Extended Data Table \ref{table:photo}), assuming reddening \cite{Schlegel_1998} (as no more recent reddening catalog covers all 23 stars), via the methodology of ref. \cite{Casagrande_2006}.
The proper motions are based on $I$ band images taken during the OGLE-IV \cite{Udalski_2015} observations of the Galactic bulge. Relative proper motions were derived from multiple epochs of data for each field \cite{Poleski_2013}, and the uncertainties are a combination of statistical and systematic (estimated to be $\sim0.4$\,mas\,yr$^{-1}$). These were converted into absolute proper motions by adding the predicted average bulge motion for each field, calculated using the Besan\c con Galaxy model \cite{Robin_2003}.  The orbits were calculated using the python package \texttt{galpy} \cite{galpy} and the galactic potential assumed in these calculations was a 3-component Milky Way-like potential \cite{Bovy_2015}. To model the uncertainty distributions, we sampled 1,000 orbits using a Monte Carlo simulation, assuming a normal distribution for the uncertainties of the input parameters. The results of this are included in Extended Data Table \ref{table:orbits}. One star, SMSS J175455.52-380339.3, has an unbound E$_{tot}$ and impractically large orbital parameters, suggesting that one or more of our input parameters need to be changed.

\subsubsection*{Code Availability}
All codes used to analyse the data presented are publicly available. In particular, the 1D LTE analysis used was made possible with the line analysis and spectrum synthesis code MOOG \cite{2012ascl.soft02009S}.

\newpage

\bibliographystyle{naturemag}
\bibliography{biblio}

\begin{thebibliography}{10}
\expandafter\ifx\csname url\endcsname\relax
  \def\url#1{\texttt{#1}}\fi
\expandafter\ifx\csname urlprefix\endcsname\relax\def\urlprefix{URL }\fi
\providecommand{\bibinfo}[2]{#2}
\providecommand{\eprint}[2][]{\url{#2}}

\bibitem{Bromm_2009}
\bibinfo{author}{Bromm, V.}, \bibinfo{author}{Yoshida, N.},
  \bibinfo{author}{Hernquist, L.} \& \bibinfo{author}{McKee, C.~F.}
\newblock \bibinfo{title}{{The formation of the first stars and galaxies.}}
\newblock \emph{\bibinfo{journal}{Nature}} \textbf{\bibinfo{volume}{459}},
  \bibinfo{pages}{49--54} (\bibinfo{year}{2009}).
\newblock \urlprefix\url{http://dx.doi.org/10.1038/nature07990}.

\bibitem{Christlieb_2002}
\bibinfo{author}{{Christlieb}, N.} \emph{et~al.}
\newblock \bibinfo{title}{{A stellar relic from the early Milky Way}}.
\newblock \emph{\bibinfo{journal}{Nature}} \textbf{\bibinfo{volume}{419}},
  \bibinfo{pages}{904--906} (\bibinfo{year}{2002}).
\newblock \urlprefix\url{http://esoads.eso.org/abs/2002Natur.419..904C}.
\newblock \eprint{astro-ph/0211274}.

\bibitem{Caffau_2011}
\bibinfo{author}{Caffau, E.} \emph{et~al.}
\newblock \bibinfo{title}{{An extremely primitive star in the Galactic halo.}}
\newblock \emph{\bibinfo{journal}{Nature}} \textbf{\bibinfo{volume}{477}},
  \bibinfo{pages}{67--69} (\bibinfo{year}{2011}).
\newblock \urlprefix\url{http://dx.doi.org/10.1038/nature10377}.

\bibitem{Keller_2014}
\bibinfo{author}{Keller, S.~C.} \emph{et~al.}
\newblock \bibinfo{title}{{A single low-energy iron-poor supernova as the
  source of metals in the star {SMSS} J031300.36-670839.3.}}
\newblock \emph{\bibinfo{journal}{Nature}} \textbf{\bibinfo{volume}{506}},
  \bibinfo{pages}{463--466} (\bibinfo{year}{2014}).
\newblock \urlprefix\url{http://dx.doi.org/10.1038/nature12990}.

\bibitem{Tumlinson_2009}
\bibinfo{author}{Tumlinson, J.}
\newblock \bibinfo{title}{{Chemical evolution in hierarchical models of cosmic
  structure. II. The formation of the Milky Way stellar halo and the
  distribution of the oldest stars.}}
\newblock \emph{\bibinfo{journal}{{Astrophys. J.}}}
  \textbf{\bibinfo{volume}{708}}, \bibinfo{pages}{1398--1418}
  (\bibinfo{year}{2010}).
\newblock \urlprefix\url{http://dx.doi.org/10.1088/0004-637X/708/2/1398}.

\bibitem{Salvadori_2010}
\bibinfo{author}{Salvadori, S.}, \bibinfo{author}{Ferrara, A.},
  \bibinfo{author}{Schneider, R.}, \bibinfo{author}{Scannapieco, E.} \&
  \bibinfo{author}{Kawata, D.}
\newblock \bibinfo{title}{{Mining the Galactic halo for very metal-poor
  stars.}}
\newblock \emph{\bibinfo{journal}{Mon. Not. R. Astron. Soc.}}
  \textbf{\bibinfo{volume}{401}}, \bibinfo{pages}{L5--L9}
  (\bibinfo{year}{2010}).
\newblock \urlprefix\url{http://dx.doi.org/10.1111/j.1745-3933.2009.00772.x}.

\bibitem{Feltzing_1999}
\bibinfo{author}{{Feltzing}, S.} \& \bibinfo{author}{{Gilmore}, G.}
\newblock \bibinfo{title}{{Age and Metallicity Gradients in the Galactic
  Bulge}}.
\newblock \emph{\bibinfo{journal}{Astrophysics and Space Science}}
  \textbf{\bibinfo{volume}{265}}, \bibinfo{pages}{337--340}
  (\bibinfo{year}{1999}).
\newblock \urlprefix\url{http://esoads.eso.org/abs/1999Ap\%26SS.265..337F}.
\newblock \eprint{astro-ph/0002123}.

\bibitem{Garc_a_P_rez_2013}
\bibinfo{author}{P{\'{e}}rez, A. E.~G.} \emph{et~al.}
\newblock \bibinfo{title}{{Very metal-poor stars in the outer Galactic bulge
  found by the APOGEE survey.}}
\newblock \emph{\bibinfo{journal}{{Astrophys. J.}}}
  \textbf{\bibinfo{volume}{767}}, \bibinfo{pages}{L9} (\bibinfo{year}{2013}).
\newblock \urlprefix\url{http://dx.doi.org/10.1088/2041-8205/767/1/L9}.

\bibitem{Howes_2014}
\bibinfo{author}{Howes, L.~M.} \emph{et~al.}
\newblock \bibinfo{title}{{The Gaia-{ESO} Survey: the most metal-poor stars in
  the Galactic bulge.}}
\newblock \emph{\bibinfo{journal}{Mon. Not. R. Astron. Soc.}}
  \textbf{\bibinfo{volume}{445}}, \bibinfo{pages}{4241--4246}
  (\bibinfo{year}{2014}).
\newblock \urlprefix\url{http://dx.doi.org/10.1093/mnras/stu1991}.

\bibitem{Keller_2007}
\bibinfo{author}{Keller, S.~C.} \emph{et~al.}
\newblock \bibinfo{title}{{The {SkyMapper} Telescope and The Southern Sky
  Survey.}}
\newblock \emph{\bibinfo{journal}{Publ. Astron. Soc. Aust}}
  \textbf{\bibinfo{volume}{24}}, \bibinfo{pages}{1--12} (\bibinfo{year}{2007}).
\newblock \urlprefix\url{http://dx.doi.org/10.1071/AS07001}.

\bibitem{Bernstein_2003}
\bibinfo{author}{Bernstein, R.}, \bibinfo{author}{Shectman, S.~A.},
  \bibinfo{author}{Gunnels, S.~M.}, \bibinfo{author}{Mochnacki, S.} \&
  \bibinfo{author}{Athey, A.~E.}
\newblock \bibinfo{title}{{MIKE: A Double Echelle Spectrograph for the Magellan
  Telescopes at Las Campanas Observatory.}}
\newblock In \bibinfo{editor}{Iye, M.} \& \bibinfo{editor}{Moorwood, A. F.~M.}
  (eds.) \emph{\bibinfo{booktitle}{Instrument Design and Performance for
  Optical/Infrared Ground-based Telescopes}} (\bibinfo{publisher}{{SPIE}-Intl
  Soc Optical Eng}, \bibinfo{year}{2003}).
\newblock \urlprefix\url{http://dx.doi.org/10.1117/12.461502}.

\bibitem{Gustafsson_2008}
\bibinfo{author}{Gustafsson, B.} \emph{et~al.}
\newblock \bibinfo{title}{{A grid of {MARCS} model atmospheres for late-type
  stars.}}
\newblock \emph{\bibinfo{journal}{Astron. Astrophys.}}
  \textbf{\bibinfo{volume}{486}}, \bibinfo{pages}{951--970}
  (\bibinfo{year}{2008}).
\newblock \urlprefix\url{http://dx.doi.org/10.1051/0004-6361:200809724}.

\bibitem{Lind_2012}
\bibinfo{author}{Lind, K.}, \bibinfo{author}{Bergemann, M.} \&
  \bibinfo{author}{Asplund, M.}
\newblock \bibinfo{title}{{Non-{LTE} line formation of Fe in late-type stars --
  II. 1D spectroscopic stellar parameters.}}
\newblock \emph{\bibinfo{journal}{Mon. Not. R. Astron. Soc.}}
  \textbf{\bibinfo{volume}{427}}, \bibinfo{pages}{50--60}
  (\bibinfo{year}{2012}).
\newblock \urlprefix\url{http://dx.doi.org/10.1111/j.1365-2966.2012.21686.x}.

\bibitem{Udalski_2015}
\bibinfo{author}{{Udalski}, A.}, \bibinfo{author}{{Szyma{\'n}ski}, M.~K.} \&
  \bibinfo{author}{{Szyma{\'n}ski}, G.}
\newblock \bibinfo{title}{{OGLE-IV: Fourth Phase of the Optical Gravitational
  Lensing Experiment}}.
\newblock \emph{\bibinfo{journal}{Acta Astronomica}}
  \textbf{\bibinfo{volume}{65}}, \bibinfo{pages}{1--138}
  (\bibinfo{year}{2015}).
\newblock \urlprefix\url{http://adsabs.harvard.edu/abs/2015AcA....65....1U}.
\newblock \eprint{1504.05966}.

\bibitem{Robin_2012}
\bibinfo{author}{{Robin}, A.~C.}, \bibinfo{author}{{Marshall}, D.~J.},
  \bibinfo{author}{{Schultheis}, M.} \& \bibinfo{author}{{Reyl{\'e}}, C.}
\newblock \bibinfo{title}{{Stellar populations in the Milky Way bulge region:
  towards solving the Galactic bulge and bar shapes using 2MASS data}}.
\newblock \emph{\bibinfo{journal}{Astron. Astrophys.}}
  \textbf{\bibinfo{volume}{538}}, \bibinfo{pages}{A106} (\bibinfo{year}{2012}).
\newblock \urlprefix\url{http://esoads.eso.org/abs/2012A\%26A...538A.106R}.
\newblock \eprint{1111.5744}.

\bibitem{Greif_2012}
\bibinfo{author}{{Greif}, T.~H.} \emph{et~al.}
\newblock \bibinfo{title}{{Formation and evolution of primordial protostellar
  systems}}.
\newblock \emph{\bibinfo{journal}{Mon. Not. R. Astron. Soc.}}
  \textbf{\bibinfo{volume}{424}}, \bibinfo{pages}{399--415}
  (\bibinfo{year}{2012}).
\newblock \urlprefix\url{http://esoads.eso.org/abs/2012MNRAS.424..399G}.

\bibitem{Placco_2014}
\bibinfo{author}{{Placco}, V.~M.}, \bibinfo{author}{{Frebel}, A.},
  \bibinfo{author}{{Beers}, T.~C.} \& \bibinfo{author}{{Stancliffe}, R.~J.}
\newblock \bibinfo{title}{{Carbon-enhanced Metal-poor Star Frequencies in the
  Galaxy: Corrections for the Effect of Evolutionary Status on Carbon
  Abundances}}.
\newblock \emph{\bibinfo{journal}{Astrophys. J.}}
  \textbf{\bibinfo{volume}{797}}, \bibinfo{pages}{21} (\bibinfo{year}{2014}).
\newblock \urlprefix\url{http://esoads.eso.org/abs/2014ApJ...797...21P}.
\newblock \eprint{1410.2223}.

\bibitem{Tumlinson_2007}
\bibinfo{author}{{Tumlinson}, J.}
\newblock \bibinfo{title}{{Carbon-Enhanced Metal-poor Stars, the Cosmic
  Microwave Background, and the Stellar Initial Mass Function in the Early
  Universe}}.
\newblock \emph{\bibinfo{journal}{Astrophys. J.}}
  \textbf{\bibinfo{volume}{664}}, \bibinfo{pages}{L63--L66}
  (\bibinfo{year}{2007}).
\newblock \urlprefix\url{http://adsabs.harvard.edu/abs/2007ApJ...664L..63T}.
\newblock \eprint{0706.2903}.

\bibitem{Schlaufman_2014}
\bibinfo{author}{Schlaufman, K.~C.} \& \bibinfo{author}{Casey, A.~R.}
\newblock \bibinfo{title}{{The best and brightest metal-poor stars.}}
\newblock \emph{\bibinfo{journal}{{Astrophys. J.}}}
  \textbf{\bibinfo{volume}{797}}, \bibinfo{pages}{13} (\bibinfo{year}{2014}).
\newblock \urlprefix\url{http://dx.doi.org/10.1088/0004-637X/797/1/13}.

\bibitem{Frebel_2007}
\bibinfo{author}{{Frebel}, A.}, \bibinfo{author}{{Johnson}, J.~L.} \&
  \bibinfo{author}{{Bromm}, V.}
\newblock \bibinfo{title}{{Probing the formation of the first low-mass stars
  with stellar archaeology}}.
\newblock \emph{\bibinfo{journal}{Mon. Not. R. Astron. Soc.}}
  \textbf{\bibinfo{volume}{380}}, \bibinfo{pages}{L40--L44}
  (\bibinfo{year}{2007}).
\newblock \urlprefix\url{http://adsabs.harvard.edu/abs/2007MNRAS.380L..40F}.
\newblock \eprint{astro-ph/0701395}.

\bibitem{Kobayashi_2014}
\bibinfo{author}{Kobayashi, C.}, \bibinfo{author}{Ishigaki, M.~N.},
  \bibinfo{author}{Tominaga, N.} \& \bibinfo{author}{Nomoto, K.}
\newblock \bibinfo{title}{The origin of low [α/fe] ratios in extremely
  metal-poor stars}.
\newblock \emph{\bibinfo{journal}{Astrophys. J.}}
  \textbf{\bibinfo{volume}{785}}, \bibinfo{pages}{L5} (\bibinfo{year}{2014}).
\newblock \urlprefix\url{http://stacks.iop.org/2041-8205/785/i=1/a=L5}.

\bibitem{Umeda_2002}
\bibinfo{author}{{Umeda}, H.} \& \bibinfo{author}{{Nomoto}, K.}
\newblock \bibinfo{title}{{Nucleosynthesis of Zinc and Iron Peak Elements in
  Population III Type II Supernovae: Comparison with Abundances of Very Metal
  Poor Halo Stars}}.
\newblock \emph{\bibinfo{journal}{Astrophys. J.}}
  \textbf{\bibinfo{volume}{565}}, \bibinfo{pages}{385--404}
  (\bibinfo{year}{2002}).
\newblock \urlprefix\url{http://adsabs.harvard.edu/abs/2002ApJ...565..385U}.
\newblock \eprint{astro-ph/0103241}.

\bibitem{Nomoto_2003}
\bibinfo{author}{{Nomoto}, K.} \emph{et~al.}
\newblock \bibinfo{title}{{Nucleosynthesis in Black-Hole-Forming Supernovae and
  Extremely Metal-Poor Stars}}.
\newblock \emph{\bibinfo{journal}{Progress of Theoretical Physics Supplement}}
  \textbf{\bibinfo{volume}{151}}, \bibinfo{pages}{44--53}
  (\bibinfo{year}{2003}).
\newblock \urlprefix\url{http://adsabs.harvard.edu/abs/2003PThPS.151...44N}.
\newblock \eprint{astro-ph/0306412}.

\bibitem{Yong_2012}
\bibinfo{author}{Yong, D.} \emph{et~al.}
\newblock \bibinfo{title}{{The most metal-poor stars. II. Chemical abundances
  of 190 metal-poor stars including 10 new stars with [Fe/H] $\leq-3.5$.}}
\newblock \emph{\bibinfo{journal}{{Astrophys. J.}}}
  \textbf{\bibinfo{volume}{762}}, \bibinfo{pages}{26} (\bibinfo{year}{2012}).
\newblock \urlprefix\url{http://dx.doi.org/10.1088/0004-637X/762/1/26}.

\bibitem{Karlsson_2013}
\bibinfo{author}{{Karlsson}, T.}, \bibinfo{author}{{Bromm}, V.} \&
  \bibinfo{author}{{Bland-Hawthorn}, J.}
\newblock \bibinfo{title}{{Pregalactic metal enrichment: The chemical
  signatures of the first stars}}.
\newblock \emph{\bibinfo{journal}{Rev. Mod. Phys.}}
  \textbf{\bibinfo{volume}{85}}, \bibinfo{pages}{809--848}
  (\bibinfo{year}{2013}).
\newblock \urlprefix\url{http://esoads.eso.org/abs/2013RvMP...85..809K}.
\newblock \eprint{1101.4024}.

\bibitem{Sharp_2006}
\bibinfo{author}{Sharp, R.} \emph{et~al.}
\newblock \bibinfo{title}{{Performance of {AAOmega}: the {AAT} multi-purpose
  fiber-fed spectrograph.}}
\newblock In \bibinfo{editor}{McLean, I.~S.} \& \bibinfo{editor}{Iye, M.}
  (eds.) \emph{\bibinfo{booktitle}{Ground-based and Airborne Instrumentation
  for Astronomy}} (\bibinfo{publisher}{{SPIE}-Intl Soc Optical Eng},
  \bibinfo{year}{2006}).
\newblock \urlprefix\url{http://dx.doi.org/10.1117/12.671022}.

\bibitem{de_Laverny_2012}
\bibinfo{author}{{de Laverny}, P.}, \bibinfo{author}{{Recio-Blanco}, A.},
  \bibinfo{author}{{Worley}, C.~C.} \& \bibinfo{author}{{Plez}, B.}
\newblock \bibinfo{title}{{The AMBRE project: A new synthetic grid of
  high-resolution FGKM stellar spectra}}.
\newblock \emph{\bibinfo{journal}{Astron. Astrophys.}}
  \textbf{\bibinfo{volume}{544}}, \bibinfo{pages}{A126} (\bibinfo{year}{2012}).
\newblock \urlprefix\url{http://adsabs.harvard.edu/abs/2012A\%26A...544A.126D}.
\newblock \eprint{1205.2270}.

\bibitem{carpy}
 \urlprefix\url{http://code.obs.carnegiescience.edu/mike}.

\bibitem{2014arXiv1405.5968C}
\bibinfo{author}{{Casey}, A.~R.}
\newblock \bibinfo{title}{{A Tale of Tidal Tails in the Milky Way.}}
\newblock \emph{\bibinfo{journal}{ArXiv e-prints}}  (\bibinfo{year}{2014}).
\newblock \urlprefix\url{http://adsabs.harvard.edu/abs/2014arXiv1405.5968C}.
\newblock \eprint{1405.5968}.

\bibitem{Barklem_2000}
\bibinfo{author}{{Barklem}, P.~S.}, \bibinfo{author}{{Piskunov}, N.} \&
  \bibinfo{author}{{O'Mara}, B.~J.}
\newblock \bibinfo{title}{{Self-broadening in Balmer line wing formation in
  stellar atmospheres.}}
\newblock \emph{\bibinfo{journal}{Astron. Astrophys.}}
  \textbf{\bibinfo{volume}{363}}, \bibinfo{pages}{1091--1105}
  (\bibinfo{year}{2000}).
\newblock \urlprefix\url{http://adsabs.harvard.edu/abs/2000A\%26A...363.1091B}.
\newblock \eprint{astro-ph/0010022}.

\bibitem{Barklem_2007}
\bibinfo{author}{{Barklem}, P.~S.}
\newblock \bibinfo{title}{{Non-LTE Balmer line formation in late-type spectra:
  effects of atomic processes involving hydrogen atoms}}.
\newblock \emph{\bibinfo{journal}{Astron. Astrophys.}}
  \textbf{\bibinfo{volume}{466}}, \bibinfo{pages}{327--337}
  (\bibinfo{year}{2007}).
\newblock \urlprefix\url{http://adsabs.harvard.edu/abs/2007A\%26A...466..327B}.
\newblock \eprint{astro-ph/0702222}.

\bibitem{Norris_2013}
\bibinfo{author}{{Norris}, J.~E.} \emph{et~al.}
\newblock \bibinfo{title}{{The Most Metal-poor Stars. I. Discovery, Data, and
  Atmospheric Parameters}}.
\newblock \emph{\bibinfo{journal}{Astrophys. J.}}
  \textbf{\bibinfo{volume}{762}}, \bibinfo{pages}{25} (\bibinfo{year}{2013}).
\newblock \urlprefix\url{http://esoads.eso.org/abs/2013ApJ...762...25N}.
\newblock \eprint{1208.2999}.

\bibitem{Asplund_2009}
\bibinfo{author}{Asplund, M.}, \bibinfo{author}{Grevesse, N.},
  \bibinfo{author}{Sauval, A.~J.} \& \bibinfo{author}{Scott, P.}
\newblock \bibinfo{title}{{The Chemical Composition of the Sun.}}
\newblock \emph{\bibinfo{journal}{Annual Review Astron. Astrophys.}}
  \textbf{\bibinfo{volume}{47}}, \bibinfo{pages}{481--522}
  (\bibinfo{year}{2009}).
\newblock
  \urlprefix\url{http://dx.doi.org/10.1146/annurev.astro.46.060407.145222}.

\bibitem{Lind_2009}
\bibinfo{author}{{Lind}, K.}, \bibinfo{author}{{Asplund}, M.} \&
  \bibinfo{author}{{Barklem}, P.~S.}
\newblock \bibinfo{title}{{Departures from LTE for neutral Li in late-type
  stars}}.
\newblock \emph{\bibinfo{journal}{Astron. Astrophys.}}
  \textbf{\bibinfo{volume}{503}}, \bibinfo{pages}{541--544}
  (\bibinfo{year}{2009}).
\newblock \urlprefix\url{http://esoads.eso.org/abs/2009A\%26A...503..541L}.
\newblock \eprint{0906.0899}.

\bibitem{Lind_2011}
\bibinfo{author}{{Lind}, K.}, \bibinfo{author}{{Asplund}, M.},
  \bibinfo{author}{{Barklem}, P.~S.} \& \bibinfo{author}{{Belyaev}, A.~K.}
\newblock \bibinfo{title}{{Non-LTE calculations for neutral Na in late-type
  stars using improved atomic data}}.
\newblock \emph{\bibinfo{journal}{Astron. Astrophys.}}
  \textbf{\bibinfo{volume}{528}}, \bibinfo{pages}{A103} (\bibinfo{year}{2011}).
\newblock \urlprefix\url{http://adsabs.harvard.edu/abs/2011A\%26A...528A.103L}.
\newblock \eprint{1102.2160}.

\bibitem{Jacobson_2015}
\bibinfo{author}{{Jacobson}, H.~R.} \emph{et~al.}
\newblock \bibinfo{title}{{High-Resolution Spectroscopic Study of Extremely
  Metal-Poor Star Candidates from the SkyMapper Survey}}.
\newblock \emph{\bibinfo{journal}{Astrophys. J. in press}}
  (\bibinfo{year}{2015}).
\newblock \urlprefix\url{http://adsabs.harvard.edu/abs/2015arXiv150403344J}.
\newblock \eprint{1504.03344}.

\bibitem{Lucatello_2006}
\bibinfo{author}{{Lucatello}, S.} \emph{et~al.}
\newblock \bibinfo{title}{{The Frequency of Carbon-enhanced Metal-poor Stars in
  the Galaxy from the HERES Sample}}.
\newblock \emph{\bibinfo{journal}{Astrophys. J.}}
  \textbf{\bibinfo{volume}{652}}, \bibinfo{pages}{L37--L40}
  (\bibinfo{year}{2006}).
\newblock \urlprefix\url{http://adsabs.harvard.edu/abs/2006ApJ...652L..37L}.
\newblock \eprint{astro-ph/0609730}.

\bibitem{Schlegel_1998}
\bibinfo{author}{{Schlegel}, D.~J.}, \bibinfo{author}{{Finkbeiner}, D.~P.} \&
  \bibinfo{author}{{Davis}, M.}
\newblock \bibinfo{title}{{Maps of Dust Infrared Emission for Use in Estimation
  of Reddening and Cosmic Microwave Background Radiation Foregrounds}}.
\newblock \emph{\bibinfo{journal}{Astrophys. J.}}
  \textbf{\bibinfo{volume}{500}}, \bibinfo{pages}{525--553}
  (\bibinfo{year}{1998}).
\newblock \urlprefix\url{http://adsabs.harvard.edu/abs/1998ApJ...500..525S}.
\newblock \eprint{astro-ph/9710327}.

\bibitem{Casagrande_2006}
\bibinfo{author}{Casagrande, L.}, \bibinfo{author}{Portinari, L.} \&
  \bibinfo{author}{Flynn, C.}
\newblock \bibinfo{title}{{Accurate fundamental parameters for lower
  main-sequence stars}}.
\newblock \emph{\bibinfo{journal}{Mon. Not. R. Astron. Soc.}}
  \textbf{\bibinfo{volume}{373}}, \bibinfo{pages}{13--44}
  (\bibinfo{year}{2006}).
\newblock \urlprefix\url{http://esoads.eso.org/abs/2006MNRAS.373...13C}.

\bibitem{Poleski_2013}
\bibinfo{author}{{Poleski}, R.} \emph{et~al.}
\newblock \bibinfo{title}{{An Asymmetric Streaming Motion in the Galactic Bulge
  X-shaped Structure Revealed by OGLE-III Proper Motions}}.
\newblock \emph{\bibinfo{journal}{Astrophys. J.}}
  \textbf{\bibinfo{volume}{776}}, \bibinfo{pages}{76} (\bibinfo{year}{2013}).
\newblock \urlprefix\url{http://adsabs.harvard.edu/abs/2013ApJ...776...76P}.
\newblock \eprint{1304.6084}.

\bibitem{Robin_2003}
\bibinfo{author}{{Robin}, A.~C.}, \bibinfo{author}{{Reyl{\'e}}, C.},
  \bibinfo{author}{{Derri{\`e}re}, S.} \& \bibinfo{author}{{Picaud}, S.}
\newblock \bibinfo{title}{{A synthetic view on structure and evolution of the
  Milky Way}}.
\newblock \emph{\bibinfo{journal}{Astron. Astrophys.}}
  \textbf{\bibinfo{volume}{409}}, \bibinfo{pages}{523--540}
  (\bibinfo{year}{2003}).
\newblock
  \urlprefix\url{http://cdsads.u-strasbg.fr/abs/2003A\%26A...409..523R}.

\bibitem{galpy}
 \urlprefix\url{http://github.com/jobovy/galpy}.

\bibitem{Bovy_2015}
\bibinfo{author}{Bovy, J.}
\newblock \bibinfo{title}{{galpy: A python library for galactic dynamics.}}
\newblock \emph{\bibinfo{journal}{{Astrophys. J.}}}
  \textbf{\bibinfo{volume}{216}}, \bibinfo{pages}{29} (\bibinfo{year}{2015}).
\newblock \urlprefix\url{http://dx.doi.org/10.1088/0067-0049/216/2/29}.

\bibitem{2012ascl.soft02009S}
\bibinfo{author}{{Sneden}, C.}, \bibinfo{author}{{Bean}, J.},
  \bibinfo{author}{{Ivans}, I.}, \bibinfo{author}{{Lucatello}, S.} \&
  \bibinfo{author}{{Sobeck}, J.}
\newblock \bibinfo{title}{{MOOG: LTE line analysis and spectrum synthesis.}}
\newblock \emph{\bibinfo{journal}{Astrophysics Source Code Library}}
  (\bibinfo{year}{2012}).
\newblock \urlprefix\url{http://adsabs.harvard.edu/abs/2012ascl.soft02009S}.
\newblock \eprint{1202.009}.

\bibitem{Ness_2014}
\bibinfo{author}{Ness, M.} \emph{et~al.}
\newblock \bibinfo{title}{{Young stars in an old bulge: a natural outcome of
  internal evolution in the Milky Way.}}
\newblock \emph{\bibinfo{journal}{{Astrophys. J.}}}
  \textbf{\bibinfo{volume}{787}}, \bibinfo{pages}{L19} (\bibinfo{year}{2014}).
\newblock \urlprefix\url{http://dx.doi.org/10.1088/2041-8205/787/2/L19}.

\bibitem{Ryde_2010}
\bibinfo{author}{{Ryde}, N.} \emph{et~al.}
\newblock \bibinfo{title}{{Chemical abundances of 11 bulge stars from
  high-resolution, near-IR spectra}}.
\newblock \emph{\bibinfo{journal}{Astron. Astrophys.}}
  \textbf{\bibinfo{volume}{509}}, \bibinfo{pages}{A20} (\bibinfo{year}{2010}).
\newblock \urlprefix\url{http://esoads.eso.org/abs/2010A\%26A...509A..20R}.
\newblock \eprint{0910.0448}.

\end{thebibliography}


\begin{addendum}
 \item[Supplementary Information] is linked to the online version of the paper at www.nature.com/nature.
 \item This paper includes data gathered with the 6.5 meter Magellan Telescopes located at Las Campanas Observatory, Chile. Australian access to the Magellan Telescopes was supported through the Collaborative Research Infrastructure Strategy of the Australian Federal Government. L.M.H. and M.A.  have been supported by the Australian Research Council (FL110100012). A.R.C. acknowledges support from the European Union FP7 programme through ERC grant number 320360. Research on metal-poor stars with SkyMapper is supported through Australian Research Council Discovery Projects grants DP120101237 and DP150103294 (PI: Da Costa). The OGLE project has received funding from the NSC, Poland,
MAESTRO grant (2014/14/A/ST9/00121 to A.U.).
 \item[Author Contributions] 
The project was initiated and lead by M.A. The photometric target selection was made by L.M.H., C.I.O., and D.M.N. using data from the SkyMapper telescope developed by B.P.S., S.C.K., G.S.D.C., M.S.B. and P.T. The low-resolution spectra were obtained by L.M.H and C.I.O.. The data were reduced and analysed by L.M.H. using software developed by A.R.C. Target selection for the high-resolution observations were done by L.M.H., M.A. and A.R.C. with the observations carried out by L.M.H. and D.Y.; the reduction and subsequent chemical analysis was completed by L.M.H. K.L. performed the non-LTE spectral line formation calculations, C.K. interpreted the observed chemical abundances in terms of supernova yields, and M.N. comparison bulge data. R.P., A.U., M.K.S, I.S., G.P., K.U., L.W., P.P., J.S., S.K., and P.M. obtained the OGLE observations, A.U. and M.K.S constructed the reference images, and R.P. measured the proper motions. The manuscript was written by M.A., L.M.H., and A.R.C. with all authors contributing comments.
 \item[Author Information] Reprints and permissions information is available at www.nature.com/reprints. Readers are welcome to comment on the online version of the paper. Correspondence and requests for materials should be addressed to L.M.H. (louise.howes@anu.edu.au). The authors declare that they have no
competing financial interests.
\end{addendum}



\newpage

\section*{}
\label{section:fig6}
\includegraphics[width=0.8\columnwidth]{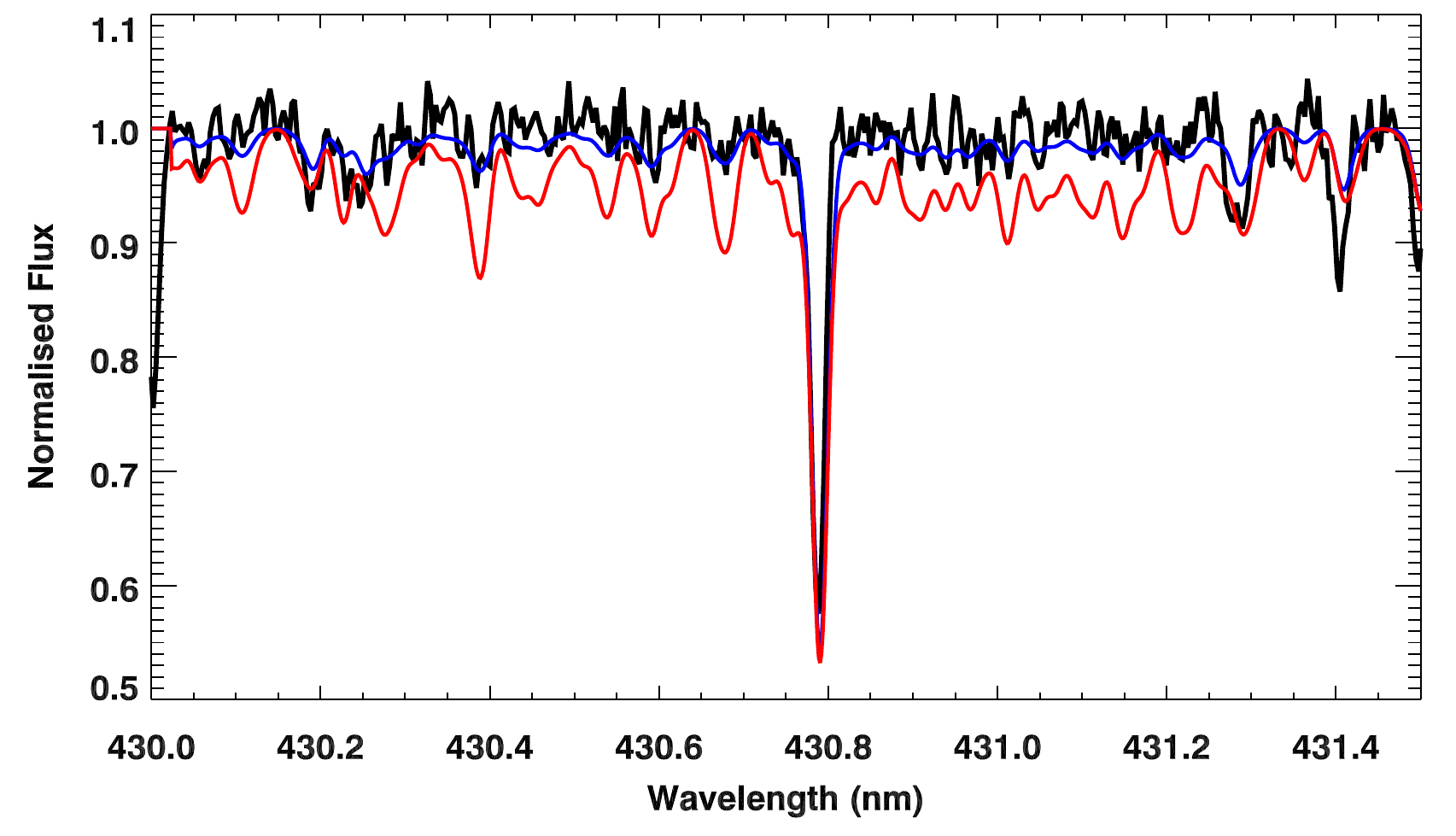}
\textbf{\\Extended Data Figure 1:} The C-H band of SMSS J181609.62-333218.7. The C-H band used to derive an upper limit for C in our most metal-poor star, SMSS J181609.62-333218.7. Synthetic spectra with abundances of [C/Fe]$=0.06$ (blue) and [C/Fe]$=0.56$ (red) are shown for comparison.

\begin{table}
\footnotesize
\caption{Coordinates and 2MASS photometry of the 23 stars observed.}
\label{table:photo}
\sffamily
\begin{tabular}{lrrrrrrr}
\hline
Name (SMSS) & RA ($^{\circ}$) & Dec ($^{\circ}$) & $l$ ($^{\circ}$) & $b$ ($^{\circ}$) & $J$ (mag) & $H$ (mag) & $K_{S}$ (mag) \\
\hline
J173823.38-145701.1 & 264.597 & -14.950 & 11.1 & 8.7 & 10.85 & 10.22 & 10.03 \\
J182048.26-273329.2 & 275.201 & -27.558 & 5.0 & -6.1 & 12.94 & 12.42 & 12.25 \\
J183744.90-280831.1 & 279.437 & -28.142 & 6.2 & -9.7 & 12.29 & 11.69 & 11.53 \\
J183647.89-274333.1 & 279.200 & -27.726 & 6.5 & -9.3 & 10.68 & 10.03 & 9.77 \\
J183812.72-270746.3 & 279.553 & -27.130 & 7.1 & -9.3 & 13.38 & 12.79 & 12.61 \\
J183719.09-262725.0 & 279.330 & -26.457 & 7.7 & -8.9 & 12.79 & 12.19 & 12.03 \\
J184201.19-302159.6 & 280.505 & -30.367 & 4.5 & -11.5 & 14.52 & 14.08 & 14.00 \\
J184656.07-292351.5 & 281.734 & -29.398 & 5.9 & -12.0 & 13.12 & 12.61 & 12.51 \\
J181406.68-313106.1 & 273.528 & -31.518 & 0.8 & -6.6 & 12.12 & 11.56 & 11.35 \\
J181317.69-343801.9 & 273.324 & -34.634 & 357.9 & -7.9 & 13.09 & 12.55 & 12.50 \\
J181219.68-343726.4 & 273.082 & -34.624 & 357.9 & -7.7 & 12.80 & 12.28 & 12.15 \\
J181609.62-333218.7 & 274.040 & -33.539 & 359.2 & -7.9 & 13.39 & 12.84 & 12.71 \\
J181634.60-340342.5 & 274.144 & -34.062 & 358.8 & -8.3 & 12.56 & 11.99 & 11.90 \\
J175544.54-392700.9 & 268.936 & -39.450 & 352.0 & -7.1 & 13.71 & 13.19 & 13.09 \\
J175455.52-380339.3 & 268.731 & -38.061 & 353.1 & -6.3 & 11.98 & 11.39 & 11.26 \\
J175746.58-384750.0 & 269.444 & -38.797 & 352.8 & -7.2 & 13.09 & 12.60 & 12.51 \\
J181736.59-391303.3 & 274.402 & -39.218 & 354.2 & -10.8 & 12.06 & 11.54 & 11.37 \\
J181505.16-385514.9 & 273.772 & -38.921 & 354.2 & -10.2 & 13.63 & 13.15 & 13.09 \\
J181921.64-381429.0 & 274.840 & -38.241 & 355.2 & -10.6 & 13.64 & 13.13 & 13.03 \\
J175722.68-411731.8 & 269.345 & -41.292 & 350.5 & -8.3 & 13.85 & 13.25 & 13.21 \\
J175021.86-414627.1 & 267.591 & -41.774 & 349.4 & -7.4 & 11.74 & 11.23 & 11.17 \\
J175636.59-403545.9 & 269.152 & -40.596 & 351.1 & -7.9 & 12.86 & 12.29 & 12.19 \\
J175433.19-411048.9 & 268.638 & -41.180 & 350.4 & -7.8 & 11.94 & 11.43 & 11.33 \\
\hline
\end{tabular}
\rmfamily
\\ RA, right ascension; Dec., declination; $l$ and $b$, Galactic longitude and latitude, respectively.
\end{table}

\begin{table}
\footnotesize
\caption{Stellar parameters of the 23 stars observed.}
\label{table:params}
\sffamily
\begin{tabular}{lrrrrrrr}
\hline
Name & $V_{helio}$  & $d_{\odot}$ & $T_{\rm eff}$  & $\log{g}$  & [Fe/H] & $\xi_{t}$  & [$\alpha$/Fe]  \\
(SMSS) & (km\,s$^{-1}$) & (kpc) & (K) & (cgs) & (dex) & (km\,s$^{-1}$) & (dex) \\
\hline
 J173823.38-145701.1 & 46.1 & 8.5 & 4599 & 0.99 & -3.36 & 2.30 & 0.12 \\
J182048.26-273329.2 & 51.5 & 6.0 & 4949 & 2.22 & -3.48 & 1.90 & 0.37 \\
J183744.90-280831.1 & -132.6 & 17.6 & 4597 & 0.98 & -2.92 & 2.05 & 0.33 \\
J183647.89-274333.1 & -381.4 & 6.6 & 4649 & 1.17 & -2.48 & 2.50 & 0.30 \\
J183812.72-270746.3 & 155.3 & 12.3 & 4873 & 1.74 & -3.22 & 1.81 & -0.01 \\
J183719.09-262725.0 & -244.7 & 10.0 & 4791 & 1.64 & -3.18 & 1.81 & 0.32 \\
J184201.19-302159.6 & 171.8 & 9.6 & 5136 & 2.55 & -2.84 & 1.96 & 0.30 \\
J184656.07-292351.5 & 91.0 & 9.5 & 4857 & 1.93 & -2.76 & 1.83 & 0.34 \\
J181406.68-313106.1 & 4.9 & 9.3 & 4821 & 1.48 & -2.82 & 1.96 & 0.22 \\
J181317.69-343801.9 & 139.3 & 6.5 & 5015 & 2.25 & -2.28 & 1.48 & 0.41 \\
J181219.68-343726.4 & -386.2 & 8.0 & 4873 & 1.94 & -2.50 & 1.93 & 0.32 \\
J181609.62-333218.7 & 27.4 & 10.4 & 4809 & 1.93 & -3.94 & 1.60 & 0.14 \\
J181634.60-340342.5 & -170.3 & 10.5 & 4821 & 1.61 & -2.46 & 1.79 & 0.06 \\
J175544.54-392700.9 & -279.6 & 13.5 & 4857 & 1.83 & -2.65 & 1.60 & 0.32 \\
J175455.52-380339.3 & 23.5 & 13.5 & 4714 & 1.10 & -3.36 & 1.80 & 0.08 \\
J175746.58-384750.0 & -59.4 & 9.1 & 5064 & 1.96 & -2.81 & 2.36 & 0.29 \\
J181736.59-391303.3 & -177.9 & 15.7 & 4612 & 1.05 & -2.59 & 2.09 & 0.32 \\
J181505.16-385514.9 & 202.1 & 5.0 & 4962 & 2.73 & -3.29 & 2.10 & 0.35 \\
J181921.64-381429.0 & -97.7 & 11.2 & 4917 & 2.02 & -2.72 & 1.94 & 0.30 \\
J175722.68-411731.8 & 63.8 & 12.4 & 4894 & 1.97 & -2.88 & 2.02 & 0.19 \\
J175021.86-414627.1 & 181.4 & 4.1 & 5015 & 2.12 & -2.60 & 1.55 & 0.30 \\
J175636.59-403545.9 & -28.8 & 9.8 & 4934 & 1.79 & -3.21 & 1.96 & 0.20 \\
J175433.19-411048.9 & -229.3 & 5.6 & 4912 & 1.91 & -3.26 & 1.94 & 0.35 \\
\hline
\end{tabular}
\rmfamily
\\ $V_{helio}$, heliocentric velocity; $d_{\odot}$, distance from the Sun to the star; $T_{\rm eff}$, effective temperature; $\log{g}$, stellar surface gravity; $\xi_{t}$, microturbulence; [$\alpha$/Fe] = ([Mg/Fe]+[Ca/Fe]+[Ti/Fe])/3. Average uncertainties: velocity, 1.0\,km\,s$^{-1}$; distance, 3.0 kpc; temperature, 160\,K; microturbulence,  0.2\,dex; log$g$, 0.14\,dex; [Fe/H], 0.09\,dex; [$\alpha$/Fe], 0.13\,dex.
\end{table}

\begin{table}
\footnotesize
\caption{Chemical abundances measured for each star (Li to Ca)}
\label{table:abunds}
\sffamily
\begin{tabular}{lrrrrrrrr}
\hline
Name (SMSS)& A(Li) & [C/Fe] & [Na/Fe] & [Mg/Fe] & [Al/Fe] & [Si/Fe] & [K/Fe] & [Ca/Fe]  \\
\hline
J173823.38-145701.1 &  & 0.49 & 0.04 & 0.17 & -0.78 & 0.27 &  & 0.12 \\
J182048.26-273329.2 &  & 0.98 & -0.28 & 0.54 & -0.63 & 0.96 &  & 0.30 \\
J183744.90-280831.1 & 0.16 & -0.20 & -0.28 & 0.44 & -0.52 & 0.58 & 0.36 & 0.25 \\
J183647.89-274333.1 &  & -0.47 & -0.24 & 0.33 & -0.66 & 0.51 &  & 0.18 \\
J183812.72-270746.3 & 0.93 & 0.22 & -0.39 & 0.05 & -1.23 & 0.14 &  & 0.03 \\
J183719.09-262725.0 &  & 0.40 & -0.19 & 0.47 & -0.77 & 0.36 & 0.41 & 0.25 \\
J184201.19-302159.6 &  & 0.34 & -0.38 & 0.26 & -0.89 & 0.38 & 0.53 & 0.37 \\
J184656.07-292351.5 & 1.04 & 0.08 & -0.30 & 0.41 & -0.95 & 0.36 & 0.58 & 0.28 \\
J181406.68-313106.1 &  & -0.51 & 0.18 & 0.23 & -0.94 & 0.32 &  & 0.16 \\
J181317.69-343801.9 & 1.05 & 0.17 & -0.33 & 0.53 & -0.82 & 0.33 & 0.63 & 0.34 \\
J181219.68-343726.4 & 1.01 & 0.19 & -0.22 & 0.30 & -0.86 & 0.25 &  & 0.31 \\
J181609.62-333218.7 &  & $<$0.06 & -0.01$<$0.91$^{a}$ & 0.20 & -1.08 & 0.54 &  & 0.00 \\
J181634.60-340342.5 &  & -0.10 & -0.53 & 0.05 & -1.08 & 0.11 & 0.21 & 0.03 \\
J175544.54-392700.9 & 0.87 & 0.12 & -0.32 & 0.29 & -0.88 & 0.34 & 0.36 & 0.29 \\
J175455.52-380339.3 &  & -0.64 &  & 0.06 & -0.88 & 0.30 &  & 0.03 \\
J175746.58-384750.0 &  & -0.04 &  & 0.37 & -1.10 & 0.44 &  & 0.24 \\
J181736.59-391303.3 &  & -0.28 & -0.11 & 0.38 & -0.69 & 0.53 & 0.53 & 0.26 \\
J181505.16-385514.9 &  & 0.23 & -0.23 & 0.21 & -0.96 & 0.15 &  & 0.36 \\
J181921.64-381429.0 & 1.04 & 0.32 & -0.24 & 0.28 & -0.82 & 0.54 & 0.44 & 0.26 \\
J175722.68-411731.8 &  & 0.42 & -0.42 & 0.21 & -0.70 & 0.48 & 0.14 & 0.13 \\
J175021.86-414627.1 & 0.98 & 0.23 & -0.37 & 0.30 & -0.82 & 0.42 &  & 0.28 \\
J175636.59-403545.9 & 0.93 & 0.65 &  & 0.30 & -0.76 & 0.45 &  & 0.11 \\
J175433.19-411048.9 & 0.92 & 0.24 & -0.03 & 0.40 & -0.74 & 0.43 & 0.44 & 0.32 \\
\hline
\end{tabular}
\rmfamily
A(Li) is the logarithmic abundance of lithium. All abundances are derived using LTE, except for Li, Na, Mg, and Ca, where non-LTE corrections have been applied. Average uncertainties: Li, 0.20; C, 0.25; Na, 0.20; Mg, 0.16; Al, 0.22; Si, 0.21; K, 0.17; Ca, 0.12.\\
$^{a}$-0.01 is the lower limit, and 0.91 is the upper limit; see Methods for details.
\end{table}

\begin{table}
\footnotesize
\caption{Chemical abundances measured for each star (Sc to Cu)}
\label{table:abunds2}
\sffamily
\begin{tabular}{lrrrrrrr}
\hline
Name (SMSS)& [Sc/Fe] & [Ti/Fe] & [Cr/Fe] & [Mn/Fe] & [Co/Fe] & [Ni/Fe] & [Cu/Fe]  \\
\hline
 J173823.38-145701.1 &  & -0.09 & -0.22 & -0.80 & 0.17 & -0.21 & $<$0.96 \\
J182048.26-273329.2 &  & 0.16 & -0.51 & -0.97 & 0.24 & -0.33 & $<$1.33 \\
J183744.90-280831.1 & 0.04 & 0.20 & -0.23 & -0.38 & 0.36 & 0.14 & $<$0.29 \\
J183647.89-274333.1 & 0.14 & 0.34 & -0.27 & -0.35 & 0.01 & 0.02 & -0.43 \\
J183812.72-270746.3 &  & -0.20 & -0.51 & -0.32 & 0.22 & -0.08 & $<$1.06 \\
J183719.09-262725.0 & 0.18 & 0.14 & -0.33 & -0.34 & 0.23 & 0.23 & $<$1.10 \\
J184201.19-302159.6 & -0.03 & 0.20 & -0.24 & -0.57 & 0.35 & -0.02 & $<$0.72 \\
J184656.07-292351.5 & 0.11 & 0.28 & -0.19 & -0.31 & 0.11 & 0.07 & $<$0.45 \\
J181406.68-313106.1 & 0.08 & 0.19 & -0.30 & -0.60 & 0.22 & 0.09 & $<$0.50 \\
J181317.69-343801.9 & 0.11 & 0.34 & -0.19 & -0.08 & 0.12 & 0.10 & $<$0.13 \\
J181219.68-343726.4 & 0.18 & 0.31 & -0.15 & -0.14 & 0.22 & 0.22 & $<$0.17 \\
J181609.62-333218.7 &  & 0.13 & -0.65 & -1.28 & 0.13 & -0.11 & $<$1.57 \\
J181634.60-340342.5 & -0.26 & 0.04 & -0.24 & -0.35 & -0.20 & -0.03 & $<$-0.05 \\
J175544.54-392700.9 & -0.05 & 0.32 & -0.22 & -0.18 & 0.24 & -0.05 & $<$0.32 \\
J175455.52-380339.3 &  & 0.02 & -0.47 & -1.05 & 0.07 & -0.12 & $<$0.85 \\
J175746.58-384750.0 & 0.12 & 0.19 & -0.44 & -0.59 & 0.09 & -0.01 & $<$0.79 \\
J181736.59-391303.3 & 0.15 & 0.24 & -0.31 & -0.34 & 0.00 & 0.08 & $<$0.10 \\
J181505.16-385514.9 & 0.40 & 0.38 & -0.54 & -0.81 & 0.17 & -0.11 & $<$1.02 \\
J181921.64-381429.0 & 0.01 & 0.32 & -0.17 & -0.44 & 0.33 & 0.25 & $<$0.50 \\
J175722.68-411731.8 & -0.22 & 0.15 & -0.31 & -0.48 & -0.04 & 0.04 & $<$0.82 \\
J175021.86-414627.1 & -0.23 & 0.26 & -0.27 & -0.31 & 0.16 & 0.16 & $<$0.27 \\
J175636.59-403545.9 & -0.28 & 0.06 & -0.29 & -0.72 & 0.17 & -0.26 & $<$0.95 \\
J175433.19-411048.9 &  & 0.22 & -0.52 & -0.91 & 0.29 & 0.02 & $<$1.08 \\
\hline
\end{tabular}
\rmfamily
\\ All abundances are derived using LTE. Average uncertainties: Sc, 0.10; Ti, 0.10; Cr, 0.21; Mn, 0.25; Co, 0.23; Ni, 0.19; Cu, 0.25.
\end{table}

\begin{table}
\footnotesize
\caption{Chemical abundances measured for each star (Zn to Eu)}
\label{table:abunds3}
\sffamily
\begin{tabular}{lrrrrrrr}
\hline
Name (SMSS)& [Zn/Fe] & [Sr/Fe] & [Y/Fe] & [Zr/Fe] & [Ba/Fe] & [La/Fe] & [Eu/Fe]  \\
\hline
J173823.38-145701.1 & 0.66 & 0.03 & 0.02 & 0.23 & -0.04 & -0.10 & \\
J182048.26-273329.2 & $<$1.01 & -0.47 &  &  & 0.03 & $<$1.33 & \\
J183744.90-280831.1 & 0.27 & -0.29 & -0.32 & 0.03 & -0.31 & $<$0.12 & \\
J183647.89-274333.1 & 0.23 & 0.18 & -0.20 & 0.45 & 0.13 & 0.17 & 0.82 \\
J183812.72-270746.3 & $<$0.79 & -1.03 &  &  & -0.70 & $<$0.77 & \\
J183719.09-262725.0 & 0.48 & 0.04 & 0.53 &  & -0.51 & $<$1.03 & \\
J184201.19-302159.6 & $<$1.15 & -0.20 & 0.04 & 0.70 & 0.16 & $<$0.94 & \\
J184656.07-292351.5 & 0.48 & -0.26 & -0.51 & 0.14 & -0.32 & $<$0.36 & \\
J181406.68-313106.1 & 0.42 & -1.61 &  &  & -0.72 & $<$0.32 & \\
J181317.69-343801.9 & 0.17 & 0.17 & -0.11 & 0.34 & 0.22 & -0.09 & 0.15 \\
J181219.68-343726.4 & 0.33 & -0.06 & -0.11 & 0.09 & 0.13 & $<$0.90 & 0.48 \\
J181609.62-333218.7 & $<$1.40 & -0.85 & 0.23 &  & $<$-0.66 & $<$1.09 & 0.91 \\
J181634.60-340342.5 & 0.21 & -0.25 & -0.65 & -0.21 & -0.32 & -0.14 & -0.11 \\
J175544.54-392700.9 & 0.36 & -0.10 & -0.2 & 0.38 & -0.11 & -0.15 & 0.21 \\
J175455.52-380339.3 & 0.63 & 0.47 & 0.01 & 0.14 & -0.57 & $<$0.66 & \\
J175746.58-384750.0 &  & -0.21 & 0.04 & 0.91 & 0.23 & $<$1.26 & 0.65 \\
J181736.59-391303.3 & 0.23 & -0.14 & -0.47 & 0.14 & -0.28 & $<$1.19 & 0.21 \\
J181505.16-385514.9 & $<$0.95 & -0.19 & 0.14 & 0.71 & 0.04 & $<$0.54 & 0.96 \\
J181921.64-381429.0 & 0.42 & -0.21 & -0.14 & 0.51 & -0.01 & 0.48 & 0.59 \\
J175722.68-411731.8 & $<$0.95 & -0.30 & -0.30 & 0.24 & -0.19 & $<$0.63 & 0.52 \\
J175021.86-414627.1 & 0.41 & -0.14 & -0.40 & 0.25 & -0.08 & $<$0.60 & 0.23 \\
J175636.59-403545.9 & 0.86 & 0.55 & 0.24 & 0.71 & -0.95 & $<$1.21 & \\
J175433.19-411048.9 & 0.50 & -0.81 & -0.29 &  & -0.42 & $<$0.91 & \\
\hline
\end{tabular}
\rmfamily
\\ All abundances are derived using LTE. Average uncertainties: Zn, 0.10; Sr, 0.20; Y, 0.12; Zr, 0.12; Ba, 0.17; La, 0.15; Eu: 0.16.
\end{table}

\begin{table}
\scriptsize
\caption{Orbital Parameters.}
\label{table:orbits}
\sffamily
\begin{tabular}{llllllll}
\hline
Name (SMSS)& $\mu_{\alpha}\cos\delta$ & $\mu_{\delta}$ & Mean $r_{peri}$ & Mean $r_{ap}$ & Mean & Mean Z$_{max}$ & E$_{tot}$\\
 & (mas\,yr$^{-1}$) & (mas\,yr$^{-1}$) & (kpc) & (kpc) & Eccentricity & (kpc) & (10$^{4}$\,km$^{2}$\,s$^{2}$) \\
\hline
J182048.26-273329.2 & -4.10 $\pm0.52$ & -6.38 $\pm0.51$ & 0.5 $^{+0.9}_{-0.2}$ & 2.9 $^{+1.7}_{-1.1}$ & 		0.70 $^{+0.11}_{-0.23}$ & 0.9 $^{+0.2}_{-0.5}$ &		-11.3 $^{+2.2}_{-1.4}$ \\
J184201.19-302159.6  & -0.38 $\pm0.90$ & -0.82 $\pm0.90$ & 1.2 $^{+1.1}_{-0.7}$ & 6.6 $^{+4.3}_{-1.6}$ & 	0.72 $^{+0.13}_{-0.20}$ & 4.3 $^{+1.7}_{-1.8}$ &		-7.1 $^{+2.1}_{-1.2}$ \\
J184656.07-292351.5 & 1.17 $\pm0.89$ & -2.32 $\pm0.89$ & 1.2 $^{+1.5}_{-0.8}$ & 4.8 $^{+2.7}_{-1.4}$ & 		0.65 $^{+0.19}_{-0.41}$ & 3.2 $^{+1.2}_{-1.4}$ &		-8.3 $^{+1.9}_{-1.2}$ \\
J181406.68-313106.1  & 2.28 $\pm0.52$ & -8.25 $\pm0.52$ & 1.1 $^{+2.3}_{-0.8}$ & 3.5 $^{+9.1}_{-1.3}$ & 		0.63 $^{+0.18}_{-0.16}$ & 2.7 $^{+8.6}_{-0.8}$ &		-9.5 $^{+5.4}_{-1.8}$ \\
J181219.68-343726.4 & -2.42 $\pm1.14$ & -1.29 $\pm1.14$ & 0.7 $^{+0.8}_{-0.4}$ & 17.3 $^{+11.3}_{-4.9}$ & 	0.92 $^{+0.05}_{-0.03}$ & 7.2 $^{+3.2}_{-1.8}$ &		-3.4 $^{+1.9}_{-1.4}$ \\
J181609.62-333218.7  & -4.14 $\pm0.64$ & -3.74 $\pm0.64$ & 1.0 $^{+3.5}_{-0.8}$ & 3.4 $^{+3.3}_{-1.8}$ & 	0.53 $^{+0.23}_{-0.28}$ & 1.9 $^{+2.6}_{-0.9}$ &		-9.9 $^{+4.0}_{-2.6}$ \\
J181634.60-340342.5  & 1.92 $\pm0.62$ & -0.31 $\pm0.62$ & 1.9 $^{+2.0}_{-1.3}$ & 7.9 $^{+8.3}_{-2.4}$ & 		0.65 $^{+0.15}_{-0.10}$ & 3.9 $^{+4.2}_{-1.7}$ &		-6.3 $^{+3.0}_{-2.0}$ \\ 
J175544.54-392700.9 & 0.03 $\pm1.49$ & -0.35 $\pm1.46$  & 1.7 $^{+2.3}_{-1.2}$ & 29.3 $^{+67.4}_{-16.5}$ & 	0.92 $^{+0.05}_{-0.04}$ & 10.8 $^{+29.6}_{-6.0}$ &		-1.4 $^{+3.7}_{-3.2}$ \\
J175455.52-380339.3  & 1.98 $\pm1.14$ & 4.76 $\pm1.14$  & 5.5 $^{+4.8}_{-3.9}$ & 811.5 $^{+1216.1}_{-796.2}$ & 0.98 $^{+0.01}_{-0.16}$ & 81.0 $^{+259.4}_{-76.6}$ &	6.9 $^{+12.2}_{-10.7}$ \\
J175746.58-384750.0 & 1.86 $\pm1.25$ & 0.17 $\pm1.25$ & 1.8 $^{+1.9}_{-1.2}$ & 5.9 $^{+5.0}_{-1.9}$ & 		0.59 $^{+0.24}_{-0.28}$ & 2.3 $^{+2.8}_{-1.1}$ &		-7.5 $^{+2.9}_{-1.8}$ \\
\hline
\end{tabular}
\rmfamily
$\mu_{\alpha}\cos\delta$ and $\mu_{\delta}$ are the proper motions in equatorial coordinates. $r_{peri}$ and $r_{ap}$ are the pericentric and apocentric radii of the orbit, respectively, Z$_{max}$ is the maximum distance the orbit reaches above/below the Galactic plane, and E$_{tot}$ is the total energy of the orbit. All values given here are the mean values from the Monte Carlo simulation of 1,000 orbits.
\end{table}

\end{document}